# Influence of roughness on capillary forces between hydrophilic surfaces

P.J. van Zwol, G. Palasantzas[*], J. Th. M. De Hosson

Department of Applied Physics, Materials Innovation Institute and Zernike Institute for Advanced Materials, University of Groningen, Nijenborgh 4, 9747 AG Groningen, The Netherlands.

Capillary forces have been measured by Atomic Force Microscopy in the plate-sphere setup between gold, borosilicate glass, GeSbTe, titanium, and UV irradiated amorphous titanium-dioxide surfaces. The force measurements were performed as a function contact time and surface roughness in the range 0.2 - 15 nm rms, and relative humidity ranging between 2 and 40 %. It is found that even for the lowest attainable relative humidity (~ 2 ± 1%) very large capillary forces are still present. The latter suggests the persistence of a nanometers thick adsorbed water layer that acts as a capillary bridge between contacting surfaces. Moreover, we found a significantly different scaling behavior of the force with rms roughness for materials with different hydrophilicity as compared to gold-gold surfaces.

*Corresponding author: g.palasantzas@rug.nl

# I. Introduction

Capillary forces [1] play a crucial role in the technology of micro/nanoelectro mechanical systems (MEMS/NEMS) since they can cause irreversible stiction between moving components [2, 3]. In general adhesive forces between surfaces have several contributions, such as van der Waals (vdW) forces, electrostatic forces, capillary forces, and other interaction forces arising from the physics and chemistry of the interacting surfaces. However, in ambient conditions, the dominant force is the capillary force associated with the formation of a nanometer size liquid capillary bridge between the two surfaces. With the advent of atomic force microscopy (AFM), force distance curves measured between surfaces and AFM probes enabled measurements of capillary adhesion forces with pN - nN sensitivity [1]. As a result further insight in its dependence on surface morphology was gained.

For accurate measurements of fundamental dispersion forces by AFM, i.e. vdW and Casimir forces at close separations (< 20 nm) using colloidal probes, capillary forces impose a substantial obstacle under ambient conditions [4]. Even relatively stiff cantilevers (4N/m) in AFM force measurements require several micrometers retraction distance (i.e., 4 micrometers for the case of smooth gold coated mica surfaces of about 0.5nm rms roughness amplitude) [5]. For softer cantilevers which may be preferred for higher force sensitivity, capillary forces fully prohibits force measurements since the cantilever would be retracted over distances far exceeding the range of the AFM piezo tube (i.e., more than 20 μm). Increasing the roughness of the films provides a remedy for reducing the capillary forces so that measurements with soft cantilevers become feasible [5]. From our previous work [5] it was concluded that the scaling of the capillary force with the rms roughness of two surfaces might depend on the wetting properties of the material. This possibility will be explored in this paper for different materials as a function of surface roughness. To this end we have extended our studies of the capillary force to phase change materials (PCM) and other materials such as borosilicate glass and titanium coatings with different roughness at different relative humidity (RH) as low as 2%. Our focus will be on the rms roughness amplitude since it has the major contribution on capillary forces as it was shown recently from theory and experiments [5, 6].

## II. Experimental procedure

Surface forces were measured with microspheres (diameter between 20 and 200 μm) mounted on cantilevers (with stiffness ~0.2 – 4 N/m) and a plate [4, 5]. The sphere must be sufficiently large to minimize the large variance often encountered in colloid probe measurements due to roughness [7]. The reverse imaging technique was employed to measure the roughness precisely at the contact area (Fig. 1 and 2) with the plate surface [8]. The gold coated spheres were imaged after performing the adhesion measurements (Fig. 1). The typical roughness for the gold films on the sphere remained unaltered indicating minimal damage when the sphere was pressed, not too firmly onto the surface. The roughness of the gold coated spheres varied between 0.8-1.5 nm rms due to local features (see Fig. 1) indicating an average value of 1.1±0.3 nm rms. In a few cases we used gold coated spheres with an increased roughness of 4 and 15 nm rms roughness value.

Careful inspection of the uncoated borosilicate spheres (Fig 2) revealed that isolated features with significant height will disappear when the sphere is pressed a few microns onto the surface (crude use), or will be deformed to the same height as the contact point of the sphere. Furthermore the roughness of the sphere on the contact area (over an area of 300 nm$^2$) decreased from 0.9 nm to 0.5 nm rms after pressing on the plate Thus, the borosilicate sphere has a roughness of 0.5 nm rms at the area of effective contact, which increases to the value of 0.9 nm rms beyond this area. As a result we will use an average of 0.7±0.2 nm rms for these spheres. Any deformation of the spheres is small and it will be taken into account in our measurements.

The spring constant of the AFM cantilevers was calibrated by either thermal tuning or electrostatic methods [4, 5]. The latter method is more precise within an error smaller than 10 %. Further, we have concluded in [5] that the contribution of electrostatic forces in addition to vdW/Casimir forces can be neglected in comparison to capillary forces in the case of relatively smooth surfaces (rms roughness < 3 nm) [5]. The attractive surface forces were measured while approaching the surface, whereas capillary adhesion was measured when retracting from the surface (pulling off) as shown in Fig. 3. For time scales below ~0.5 s the vdW/Casimir force (plus an electrostatic due to contact potential of ~200 mV) is visible when approaching the surface. Notably the forces needed to release the sphere from the surface are enormous: the cantilever bends approximately 10 nm due to the vdW/Casimir force (possibly more but we cannot measure this due to jump-to-contact), while capillary forces result in bending of 1.5 μm during retraction, which is more than three orders of magnitude larger than that from other surface forces [5].

Finally, we would like to point out that during contact of the sphere with the plate the load is kept low allowing the cantilever to bend only less than 10 nm inwards. For these loads, the measured pull off forces were largely invariant (indicated by a lower standard deviation in Fis. 4 and 5) for surfaces in the smooth or rough limits. In the intermediate roughness regime and for titanium plate surfaces where the force was increased by a substantial factor after repeating the measurement a few times at the same place, the statistical deviations appear larger. This may indicate some deformation of the roughness on the plate or the collection of water at the specific contact point of the two surfaces i.e. local alteration of the water film due to droplet formation [9]. The force measurement is performed at different locations on the plate surface, and sufficient statistical sampling is performed.

### III. Analysis of the force measurements

Since the pull off force we measure is a combination of both preadsorbed water layers and capillary condensation, we first describe the effect of contact time on capillary forces. The typical contact time used in this paper is about 10 ms (for Figs. 5 and 6). We have also tested much longer contact times up to 100 seconds. This is shown in Fig. 4 for the case of gold surfaces. Other surfaces were tested as well and similar behavior was found. For the smooth regime the capillary force increased about a factor of two or less, which is in agreement with results in [10]. For the intermediate (roughness) regime the force can increase up to a factor of five. For the rough regime no increase in the force was measured at all. Thus, especially in the intermediate regime, the force tends to increase with repeating the measurement [4] and increasing the contact time. Over time the water layer wets better in the pores of a rough surface. The effect is reversible and as a result when the contact time is decreased from 100s to 10ms the pull off force was decreased again. Although the typical time for capillary condensation is less than 5ms, stabilization of the bridge may be reached only after a second for larger bridges (radius ~400nm) [10]. Since our system is over 100 times larger (sphere radius), it is intriguing to see that the stabilization time is in the order of 100 seconds (or more) in the case of smoother surfaces. The reason that we do not see any increase for the rough regimes can also be explained by the fact that only the aspherities (correlation length ~50 nm) interact through capillary bridges (inset, Fig. 5), for which we expect a much smaller stabilization time (<5ms) [10].

Adhesion measurements between gold-gold surfaces in the past indicated no significant influence of the RH on the measured forces in the range between 15 and 60% [5]. Here we performed similar measurements for 2≤RH≤40% indicating again the absence of any influence as Figs. 5 and 6 indicate. Since our samples have been exposed to ambient conditions during transfer from the vacuum evaporator ($10^{-6}$ mbar) to the AFM, adsorbed water layers will form on the exposed surfaces (both on plate and sphere). These findings indicate that when these layers form they will persist even at very low RH close to 1 % (even upon constant exposure for one day). Therefore, the adsorbed water layers must be thicker than $2R_k$ (maximum separation distance over which capilary condensation occurs [1]) with $R_k$ given by the Kelvin equation $R_k = -(\gamma V_m / RT)[\log(RH)]^{-1}$ with $\gamma$ the liquid surface tension and $V_m$ the molar liquid volume [1]. Indeed, for water with T=300 K and $\gamma$=73 mJ/m² we have $(\gamma V_m / RT) \approx 0.54$ nm [1] resulting in $2R_k$~0.54 – 2.7 nm for RH=1 – 40 %.

Measurement of the contact angles (inset of Fig. 6) yields for gold 77±7°, for titanium 72±3° and for PCM 54±3°. The contact angles reported in literature are 70° for gold [11, 12], 80° for evaporated titanium (which oxidizes) [13], and 30° for the hydrophilic borosilicate glass surfaces (>80% $SiO_2$) [14]. The measured forces are higher for the rougher more hydrophilic materials. Indeed, for the relatively smooth surfaces, the theory gives a capillary force (ignoring weak vdW interactions) [17]

$$F_c = \pi \gamma \tilde{R} \{-\sin\varphi + [\cos(\theta_1+\varphi)+\cos(\theta_2)]\sin^2\varphi(1-\cos\varphi+D/\tilde{R})^{-1}\} \\ + 2\pi\gamma\tilde{R}\sin\varphi\sin(\theta_1+\varphi) \quad (1)$$

with $\tilde{R}$ the local sphere radius and $\theta_{1,2}$ the contact angles on the two surfaces (sphere-plate), D the closest distance between sphere and plate surface, and $\varphi$ the filling angle (see right inset in Fig. 3) [17]. For macroscopic objects or D<<$\tilde{R}$ we have $\varphi \approx 0$ and Eq.(1) yields the capillary force independent of RH [17]

$$F_c \approx 2\pi\gamma\tilde{R}[\cos(\theta_1)+\cos(\theta_2)]. \qquad (2)$$

The latter clearly indicates that the force increases with decreasing contact angle θ or increasing hydrophilicity. We should point out that in Eqs.(1)-(2) $\tilde{R}$ represents the actual sphere radius (10 – 50 μm) for smooth sphere-plate surfaces (smooth limit), while for rough surfaces where contact down to a single or few asperities is possible (leading to formation of nanosize capillary bridges) $\tilde{R}$ (i.e., ~50 nm) represents an average asperity radius (rough limit) [5, 17].

For Atomic Layer Deposited (ALD) 56 nm thick titanium dioxide ($TiO_2$) film (amorphous) the contact angle was 80±5º prior and 6±3º after UV irradiation overnight. The adhesion force measurements for the $TiO_2$ surfaces were performed within a few minutes after UV illumination since the contact angle increases rapidly (~20º after 30 minutes). However, no appreciable increase in the force took place beyond the measurement uncertainty, indicating that UV induced super hydrophilicity alone cannot drastically increase the force. At least these measurements indicate that the adsorbed water layer on the plate does not increase upon UV irradiation. Nevertheless, as it emerges from Fig. 4 and 5 the adhesive forces remain large for the other more hydrophilic rougher surfaces.

Earlier studies have shown that the contact angle hardly changed when the rms roughness increased from 2 to 5 nm [13]. We did not measure any difference in contact angle for gold with rms roughness between 1.5 – 9 nm . Therefore, since borosilicate glass followed by PCM has the lowest contact angle, the increased hydrophilicity can explain the profound difference in scaling with roughness compared to the less hydrophilic materials. On the other hand, in the smooth surface limit, the measured forces correspond qualitatively to the theoretical predictions giving higher forces for the materials with lower contact angles when comparing Ti to GeSbTe materials (fig 6), but not for borosilicate compared to gold (fig 5). However the deviations are relatively small and the thickness off the adsorbed layers (ignored in equation 2) also play a role.

The measured capillary forces can be related to the height of the capillary bridge between the rough contacting surfaces. The idea is that the water bridge is closed (i.e. wets the sphere and plate surface) if it is higher than the highest roughness peaks (Fig. 5 inset) or approximately 3.5 times the rms roughness (we define the thickness of a water film from the mean height of a rough film), while if the water layer is thinner the force drops rapidly. For gold the adhesive force drops to 1/10 of the maximum measured force for the smoothest sample approximately at 3 nm rms. Since in this case the highest asperities defining the contact distance between sphere and plate surface $d_o$ (~3.5x3 nm) of roughly 10 nm [4], indicating a water layer film thickness off at most $d_0/2$~5 nm. Thus if $d_o$ is the distance upon contact due to roughness (where the adhesive force drops by 1/10 from its maximum value) and $d_{sph}$ the thickness of the water film on the sphere surface, the estimated water layer thickness $d_w$ on the rough plate surface is given $d_w \approx (d_o - d_{sph})$. For the PCM material a value 1/10 is only reached above 6 nm rms giving a capillary bridge height approximately 21 nm, and therefore for the apparent adsorbed water film on PCM we obtain a water layer thickness of $d_w \sim 16$ nm using a water layer thickness of ~5 nm for gold coating on the sphere surface. Using the same procedure we obtain for the adsorbed water layer on borosilicate glass $d_w \sim 15$ nm. The latter values are consistent with the fact that both surfaces wet more easily than gold. Titanium and $TiO_2$ show roughly the same behavior as gold. Therefore from our measurements we conclude that the thickness of the water layer on (oxidized) Ti or (UV irradiated) $TiO_2$ must be no more than ~5nm.

More evidence on the thickness of the bridge can be found from the force distance curves. In [4] we concluded that the jump to contact of 5nm occurring at 12nm separations was due to the capillary bridge. When using a stiffer cantilever with a 5 times smaller gold coated sphere this jump to contact remained the same, being still 5nm. Thus apparently nanometers thick water droplets reside on the top op the asperities pulling the sphere into contact with the plate, as vdW forces alone cannot lead to such jump to contact. This limits the distance at which one surface can approach the other to above roughly 10 nm [4].

Scanning tunneling microscopy (STM) studies on the measurement of adsorbed water layers [11, 15] provide further insight to our findings for the water layer thickness. Some STM studies reported adsorbed water layers of several Angstroms up to tens of nm thick on mica [15], gold, hydrophobic graphite, or hydrophilic (oxidized) titanium films even for RH as low as 15% [11]. For water on mica, the reverse process, i.e. evaporation of the adsorbed water layer when lowering RH, was also observed [15].

Finally, the elasticity of the materials used for force measurement may have major impact on capillary forces. Elastic materials with low Young's modulus deform more easily resulting in larger contact areas between rough surfaces during deformation [6]. Contrary to permanent plastic deformation, elastic deformation is more difficult to quantify. The Young's modulus E of the materials used here (ceramics or metals) are quite high, with values of 72, 78 and 116 GPa for borosilicate glass, gold and titanium, respectively. As Fig. 4 indicates, since we measured the largest adhesive forces for the stiffer material (keeping the contact loads low during measurement), we can conclude that elastic deformation plays a negligibly small role in the present measurements.

## IV. Comparison with other studies and discussion

For rough gold surfaces covered with hydrophobic and hydrophilic self assembled monolayers the pull off forces increased only by a factor of two between 0-100% RH, and only by 20% in the range 0-40% RH [16], which is close to our statistical measurement error. In addition, between hydrophilic and hydrophobic surfaces the force difference found by measurement was only a factor of two [16, 18]. These differences are in fact substantially smaller than the factor of ~100 decrease in the measured force when the surface roughness increases a few nm. In ref. [17] a decrease in the force of only a factor of five was measured with increasing roughness. This may be explained by that either the rough limit (roughness asperity interaction leading to nanocapillary formations) or the smooth limit (where the whole surface of sphere and plate interact through the capillary force) was not reached. A smaller sphere-diameter/roughness-asperity ratio may also lead to a smaller difference between the two limits. Although we have used 20 μm borosilicate spheres and 100 μm gold coated spheres, we do not observe a difference between the two concerning force limits (Fig. 5). This can be attributed to multiple asperity water bridge formation. Simply less asperities are involved in water bridge formation for the smaller sphere in case of a rough plate.

It is worth pointing out that the forces measured here in the rough limit correspond in magnitude to those with normal AFM tips. The radius of the tip (i.e., 50 nm) has the same dimensions as roughness asperities ($\tilde{R}$ ~50 nm) of our samples, indicating that our results are consistent with these studies for rough surfaces [21]. The possibility of multiple asperity water nanobridge formation is very likely to occur as stated before (inset in Fig. 5).

In a MEMS study [19], where the sample was free of adsorbed layers when the initial measurements were performed at 0% RH, revealed only large capillary forces for RH above 60%. For RH lower than 60% only vdW adhesion was measured indicating the absence of adsorbed water. For RH > 60 % the Kelvin equation (described earlier in sec. III) predicts rapidly growing water layers on surfaces. Although the polysilicon surfaces used in [19] are somewhat rougher than ours, the question remains whether the reverse process works as well. In some studies this appears to be the case [14, 18], whereas we do not measure any large changes (outside the error bars) when the RH is decreased to almost 0%. Indeed, in [20] it was shown that the capillary force when measured between gold surfaces decreased by only less than a factor of two when 'drying' the samples for 68 hours in dry nitrogen (leading to < 1 % RH) indicating still the presence of water meniscus in agreement with our findings. Only in UHV conditions the capillary force vanished leaving a much weaker remnant adhesion due to vdW interactions.

If we compare the height of the capillary bridges between gold-gold surfaces with those estimated by STM studies [11], the estimated water layer agree in both cases of having a thickness ~5 nm. For titanium, which oxidizes, a large value of 50 nm was found by STM [11]. In contrast our studies indicate that oxidized titanium is less hydrophilic than gold, and the force has not increased for rougher surfaces compared to gold. Although we would expect large adhesive forces even at 10 nm rms roughness for titanium if the water layer or capillary bridge was 50 nm high as predicted by STM [11], the force drops fast at 3 nm rms indicating a water layer thickness of at most 5 nm or less.

Similarly, the UV irradiated super hydrophilic $TiO_2$ surface did not show any sign of very thick (> 10 nm) adsorbed water layers (or high capillary bridges) either. Moreover, in [21] X-ray studies provided evidence that the adsorbed water layer on $TiO_2$ is only 1 nm thick, indicating a discrepancy with our results while the discrepancy with the STM results is very large (more than an order of magnitude) [11]. On the other hand the work in ref. [21] shows that the thickness of the water layer on $TiO_2$ hardly changed upon UV irradiation, which is in agreement with our results as we did not measure any change for the adhesive force.

Note that ellipsometry studies on gold surfaces revealed also that for a free gold surface the adsorbed water layer appears to be smaller than 1 nm [22]. This leads us to conclude that the effective water layer thickness, in the form of a capillary bridge can be larger than the thickness of a water layer on a free surface. Capillary bridges can grow over time and wet better in the pores of a rough surface. This wetting improves even more for more hydrophilic surfaces. Thus the adsorbed water layer on a free surface is increased when capillary condensation takes place. Note also that since the jump to contact remained 5nm for gold or glass surfaces independent of the cantilever stiffness, if we subtract from this value the value due to capillary condensation (2 $R_k$ = ~2nm) then we obtain a value of 1.5 nm thick for each individual free gold surface. This is in good agreement with the results in [21, 22].

### V. Conclusion

In summary, we have measured large pull off forces at very low relative humidity for different materials due to presence of significant adsorbed water layers (capillary bridges) between the contacting surfaces. The measured capillary forces did not change when decreasing RH from 40 to 2 % indicating persistent nanometer thick adsorbed water layers.

Moreover, different scaling behavior of the force with rms roughness for different materials was found depending on the degree of hydrophilicity, which was quantified by contact angle measurements. The apparent height of the water bridge as estimated from the scaling of the force with roughness, which appears to be larger for the more hydrophilic surfaces, is larger than the thickness of adsorbed water on free surfaces indicating better pore wetting for more hydrophilic surfaces. Jump to contact measurements with different stiff cantilevers provide more insight in the thickness of an adsorbed water layer on a free surface. Once the effect of capillary condensation is subtracted, the obtained value for the thickness of adsorbed water on a free surface is in good agreement with non-contact studies such as ellipsometry and x-ray measurements.

**Acknowledgements**

The research was carried out under project number MC3.05242 in the framework of the Strategic Research programme of the Materials Innovation Institute (M2I, the former Netherlands Institute for Metals Research or NIMR). Financial support from the M2I is gratefully acknowledged. We are grateful to R. van de Sanden and W. Keuning for supplying the ALD TiO$_2$ film, P. Rudolf and T. Fernandez for the contact angle and UV measurements, G. Krishnan for help with the deposition system and B.N. J. Persson for useful discussions.

**Figure captions**

**Figure 1 (a)** (Color online) a clean (no peaks) 20 micron borosilicate sphere (0.7nm rms) and **(b)** a 100 micron Au coated sphere (1 nm rms). The gold sphere was imaged after it was used in Casimir/capillary force measurements (with a 4N/m cantilever). **(c)** Inverse imaging grating (nt-mdt tgt 1). **(d)** A flattened zoom where the topography of the gold sphere is shown. The typical roughness of a gold film is clearly visible (6nm top to bottom at the centre with peaks of 10nm height farther away from the centre) indicating little damage to the film.

**Figure 2 (a)** (Color online) Images of a 20 μm borosilicate sphere before and **(b)** after performing capillary force measurements. This sphere was pressed on the surface a few times (a few microns with a 0.4N/m cantilever) and the roughness at the contact point decreased somewhat from 0.9 to 0.5nm rms (4.5 to 2.5nm top to bottom). Furthermore isolated larger features farther away from the contact point will deform and have the same height as the contact point (see the profiles). The larger feature appears to have moved laterally somewhat after measurement but this is due to the whole cantilever and sphere making a slightly different angle with the grating/surface. Small spheres are much more susceptible to contact deformation than the larger ones due to the much smaller contact area.

**Figure 3** (Color online) Approach and retract curves set out in time. vdW/Casimir forces (left inset) are much weaker than the capillary pull off forces for smooth surfaces. Right inset shows the capillary meniscus diagram for Eq. (1).

**Figure 4** (Color online) Pull off measurements as a function of contact time for a 100 μm gold coated sphere and different rough gold surfaces (RH 60%). In the smooth regime (total roughness of both surfaces – sphere + surface: 2.5 nm rms) (○), intermediate regime (total roughness: 3.5 nm rms) (□), and the rough regime (total roughness: 5.7 nm rms or more) (Δ).

**Figure 5** (Color online) Force measurements at 40%RH for gold plates with different roughness and a 17.3±1.4μm borosilicate glass sphere (■) compared to a 99.5±1.5μm gold coated polysterene sphere (□). The more hydrophilic borosilicate surface gives rise to a much increased force for the rougher surfaces indicating a thicker adsorbed water layer. The inset shows capillary adhesion in the smooth regime (complete wetting) and the rough regime (asperity wetting). The theoretical values for gold-gold (dashed line) and glass-gold (solid line) (contact angle 70º for gold and 30º for glass) are also shown for the smooth and rough limit (roughness asperities are typically ~$10^3$ times smaller than a microsphere with correlation length ~50nm).

**Figure 6** (Color online) Force measurements between a 100μm gold coated polysterene sphere and gold (○), Ti (◊), GeSbTe (□ (rougher plates) and Δ (rougher spheres)) and $TiO_2$ (►) plates. In some cases spheres with different roughness were used. The inset shows the contact angles of all the materials. We found larger forces for the rougher GeSbTe films which are also the more hydrophilic films indicating thicker adsorbed water layers. The forces remain the same in the regime 2%-40%RH. For the $TiO_2$ films we did also not notice large differences after UV illumination. The theory for gold-gold (horizontal lines) is also shown for the smooth and rough limit using Eq. (2).

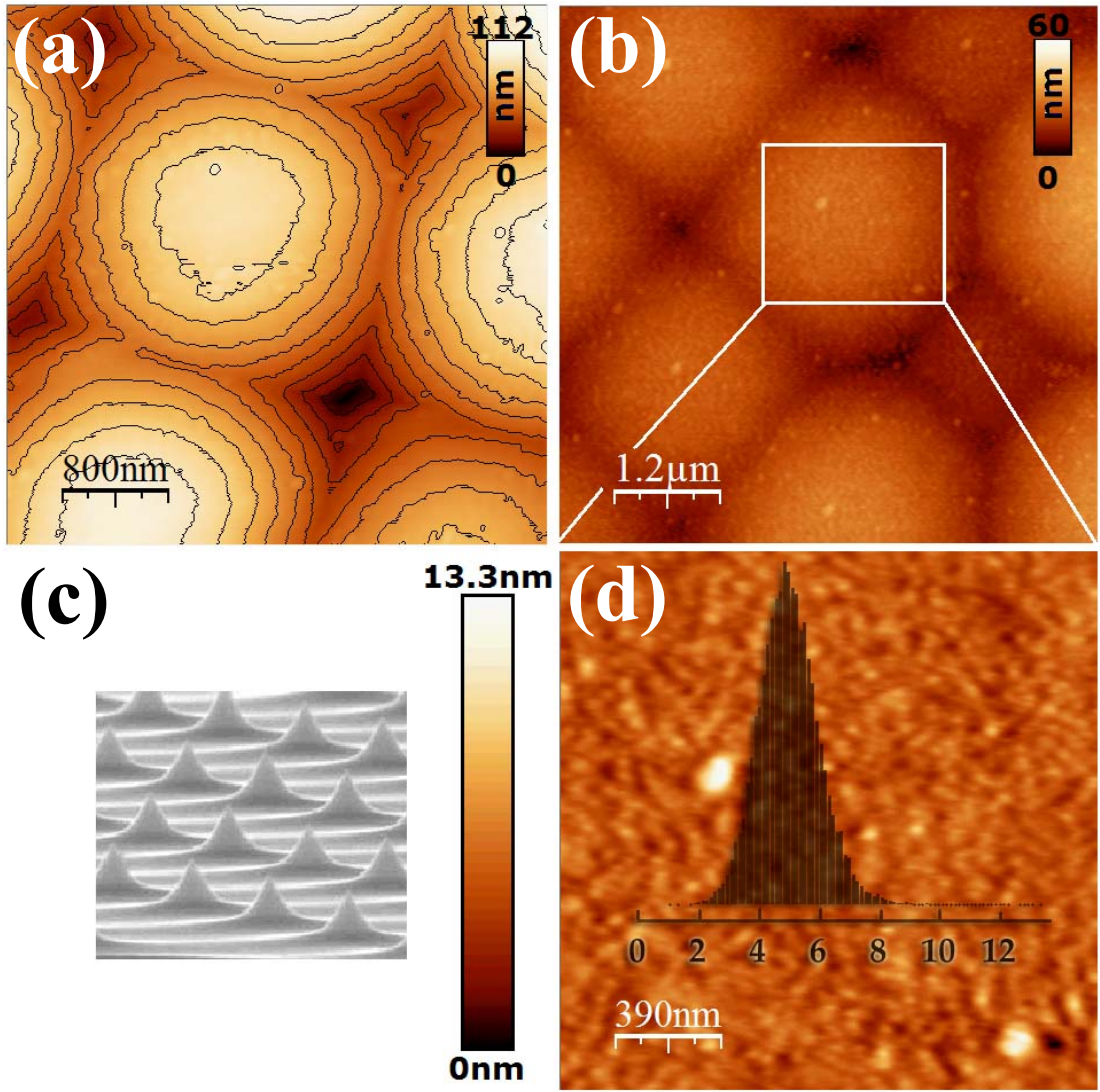

**FIGURE 1 (EU 10289)**

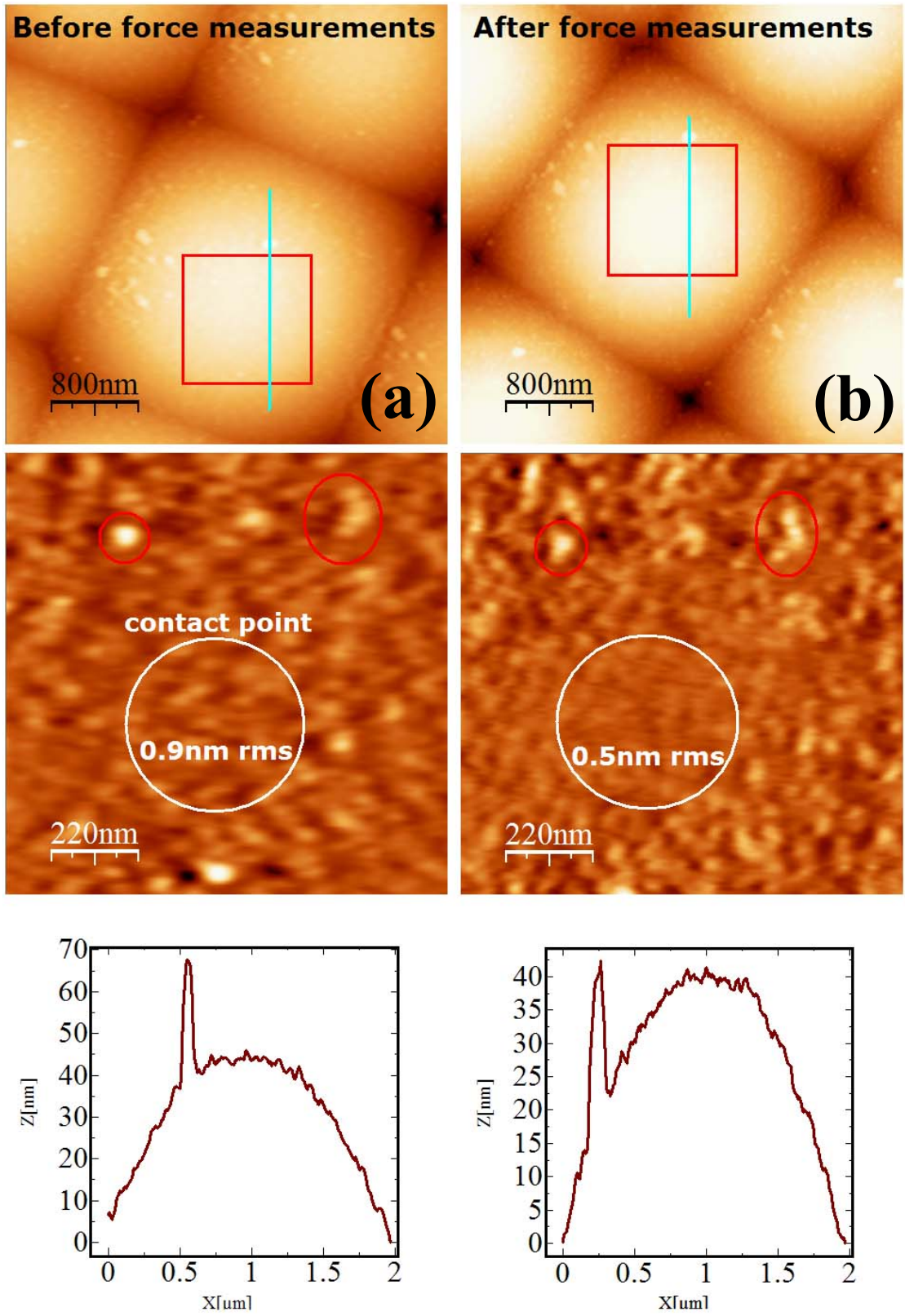

**FIGURE 2 (EU 10289)**

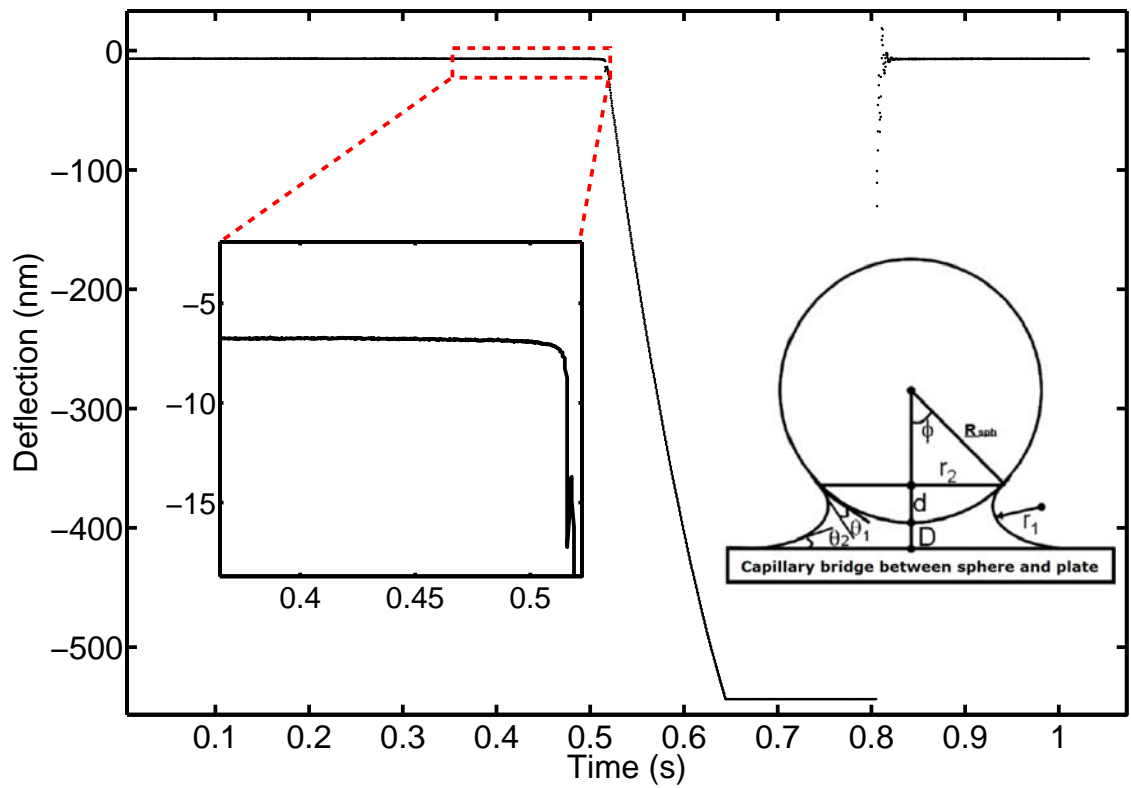

**FIGURE 3 (EU 10289)**

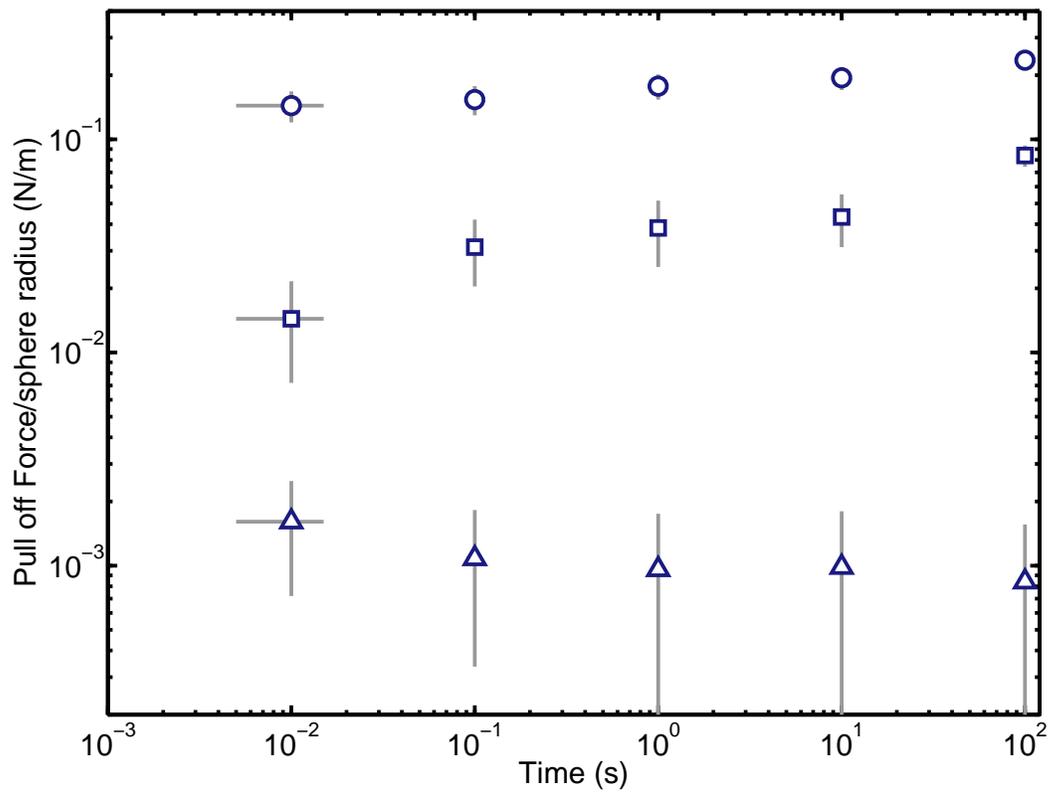

**FIGURE 4 (EU 10289)**

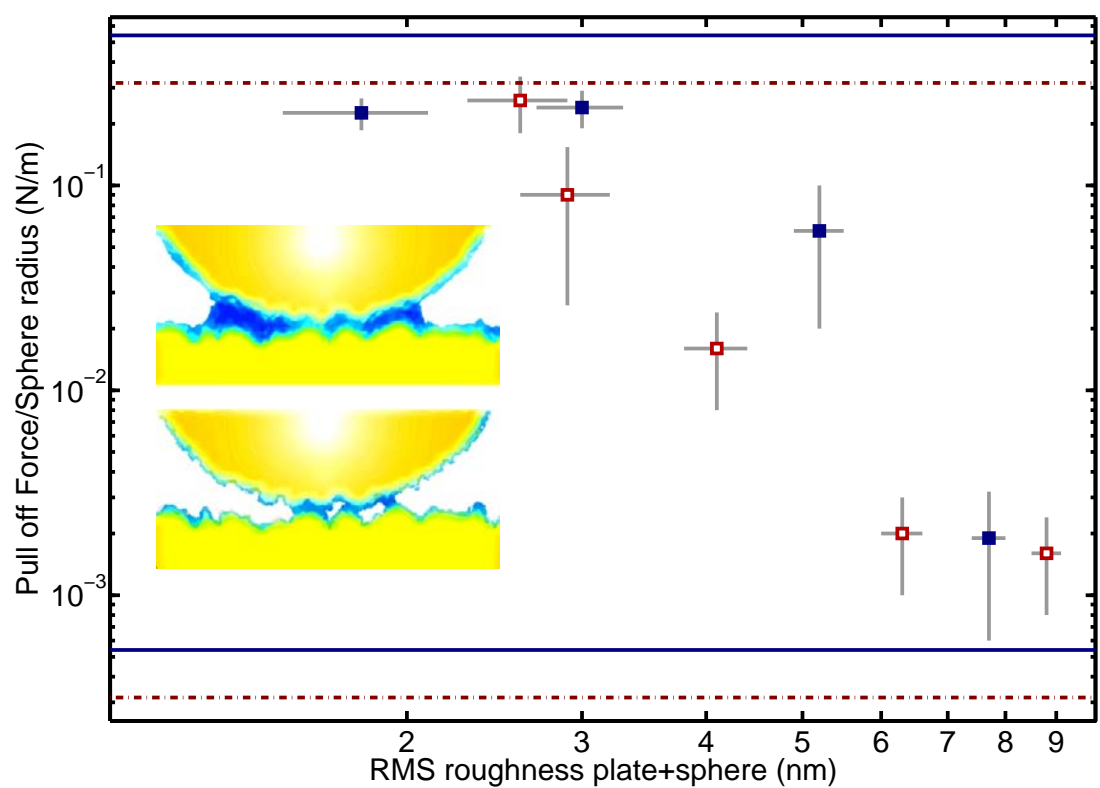

**FIGURE 5 (EU 10289)**

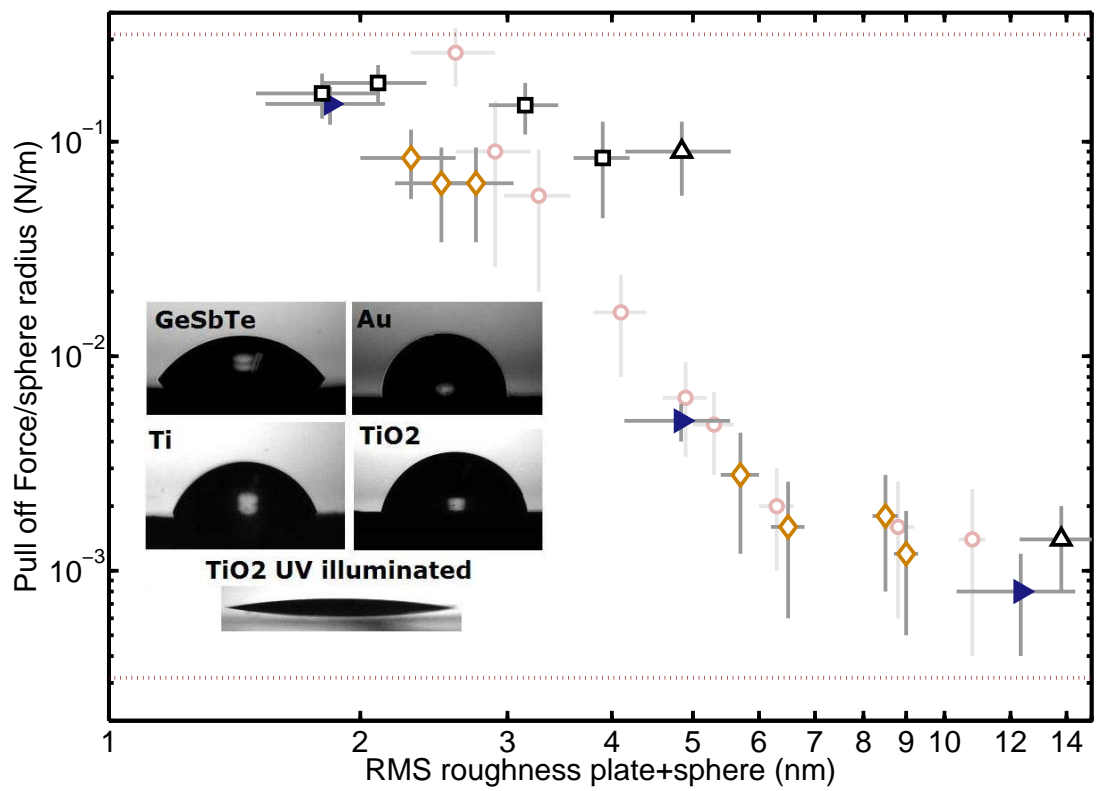

**FIGURE 6 (EU 10289)**